# 170 - 260 GHz Sub-THz Optical Heterodyne Analog Radio-over-Fiber Link for 6G Wireless System

Amol Delmade[1*], Alison Kearney[2,1], Simon Nellen[3], Robert B. Kohlhaas[3], Colm Browning[1, 4], Martin Schell[3], Frank Smyth[2] and Liam Barry[1]

(1) School of Electronic Engineering, Dublin City University, Ireland *amol.delmade2@mail.dcu.ie
(2) Pilot Photonics Ltd., Invent Centre, Dublin City University, Glasnevin, Ireland
(3) Fraunhofer Institute for Telecommunications, Heinrich Hertz Institute (HHI), Berlin, Germany
(4) mBryonics Ltd., Unit 13 Fiontarlann Teo, Westside Enterprise Park, Galway, H91 XK22, Ireland

**Abstract** *The frequency and phase fluctuations of free-running lasers limit the performance of optical heterodyne sub-THz systems – especially for low subcarrier spacing OFDM signals. Digital impairment compensation is implemented here for the successful generation of 170 - 260 GHz sub-THz OFDM signals over 10km analog-RoF link. ©2023*

## 1. Introduction

Data transmission in the sub-THz (100 – 300 GHz) and THz (300 GHz – 3 THz) frequency bands will be an important aspect of 6G wireless systems to achieve more than 20 and 100 Gb/s data rates, respectively [1]. While the shift towards the use of high-frequency carriers will help in terms of increased channel bandwidth, higher directivity and small antenna aperture, multiple issues with link implementation such as higher transmission loss and efficient techniques for the generation and distribution of sub-THz/THz signals need to be addressed [1].

Dense deployment of antenna sites and efficient optical fronthaul network, to connect with the central office (CO), is required to increase the coverage of 6G systems [2]. Analog radio-over-fiber (A-RoF) transmission has shown promising results for efficient fronthaul implementation as it preserves the spectral efficiency of the wireless data and avoids the use of digital-to-analog converters at the remote antenna sites [3-5] and the same is considered here.

The frequency multiplier-based electronic techniques currently used for sub-THz signal generation suffer from higher conversion loss and increased phase noise [6–7] – hindering their wide deployment. An alternate solution based on optical heterodyning, wherein the beating of two optical carriers with a spacing equal to the desired carrier frequency on a high-speed photodetector (PD), is being considered for the millimetre, sub-THz and THz signal generation [7-9]. This technique can be easily integrated with the A-RoF link to achieve seamless generation and distribution of sub-THz signals. An optical heterodyne A-RoF link is demonstrated here for 170 - 260 GHz signal generation.

Tunable lasers are commercially available to achieve the desired sub-THz/THz frequency separation, however there are very few suppliers for high-efficiency PDs working at such high frequencies. We used two free-running tunable fiber lasers from NKT photonics and a waveguide integrated PD antenna (WIN-PDA) from HHI [10] for this demonstration. A receiver containing a horn antenna and multiplication stage was used for frequency down-conversion from sub-THz to intermediate frequency (IF).

Free-running tunable lasers exhibit frequency drift, due to temperature fluctuations, which will lead to frequency offset (FO) with respect to the desired sub-THz frequency upon heterodyning. Also, the phase noise (PN) of the beating lasers can degrade the system's performance – especially for low subcarrier (SC) spacing orthogonal frequency division multiplexing (OFDM) signals synonymous with the 5G [11]. It is necessary to compensate the FO and PN for successful A-RoF system implementation. In this work, digital signal processing (DSP) compensation based on the Schmidl & Cox algorithm [12] and pilot tones is implemented for FO & PN impairment mitigation, respectively, for low SC spacing (2–40 MHz) OFDM signals. Various reported demonstrations [13-16] employ transmission of either single carrier waveforms or high SC spacing (> 100 MHz) OFDM signals for which the effect of FO and PN is minimal. The successful transmission of the low SC spacing OFDM signals at the sub-THz frequency range will be an important aspect for certain 6G applications and is demonstrated here.

Initially, signal generation at 200 GHz frequency is demonstrated for variable signal bandwidth (BW) from 200 MHz to 4 GHz. Subsequently, the frequency of sub-THz signals is varied from 170 to 260 GHz by tuning the spacing between the fiber lasers.

## 2. Experimental Details

The schematic of the experimental testbed is shown in Fig. 1. At the CO, two free-running fiber lasers with an initial frequency separation of 195 GHz ($\Delta F_c$) were used. The optical carrier from laser 1 (orange carrier) was modulated with an OFDM data signal. OFDM signals were generated at 5 GHz IF ($F_{IF}$) in Matlab and then converted to the electrical domain using an arbitrary waveform generator (AWG). The OFDM signals BW was varied from 200 MHz to 4 GHz by changing the 64-QAM data subcarriers spacing from 2 to 40 MHz. It resulted in raw data rates of 1.2 Gb/s to 24 Gb/s, respectively.

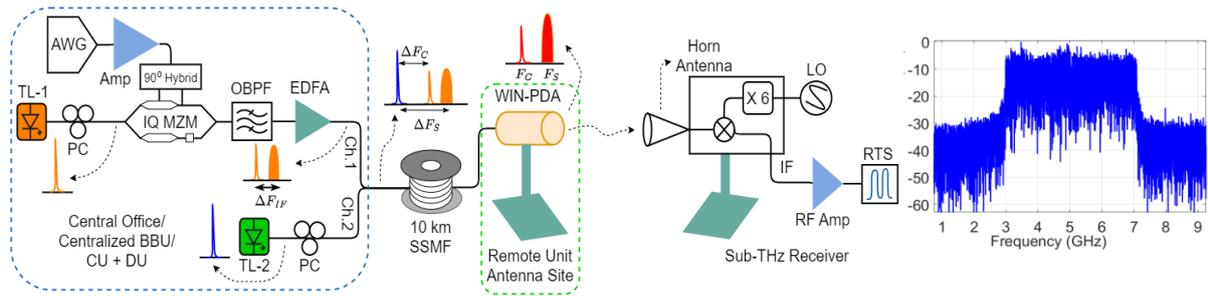

**Fig. 1**: Optical heterodyne A-RoF experimental setup including figurative optical spectrums at various stages. The inset shows the received 4 GHz bandwidth sub-THz OFDM signals spectrum after frequency down-conversion to IF.

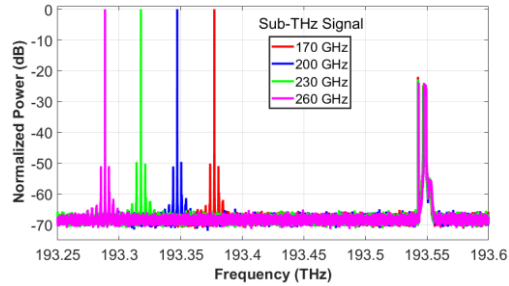

**Fig. 2:** Optical spectrum of the combined signal for different frequency Sub-THz signals generation cases.

In order to avoid the fiber induced dispersive fading, an optical single sideband (OSSB) data signal was generated using an IQ Mach Zehnder Modulator (IQ-MZM) and $90^0$ electrical hybrid as shown in Fig. 1. The bias of IQ-MZM was adjusted to reduce the power of the carrier. An optical band pass filter (OBPF) was used to filter the residual lower data sideband. An Erbium-doped fiber amplifier (EDFA) was used to boost the power of the OSSB signal before combining it with the second optical carrier (blue carrier) 195 GHz away. The combined signals spectrum is shown in Fig. 2 along with other cases.

The combined signal consisting of two optical carriers, separated by 195 GHz, and a data signal 5 GHz away from one carrier was transmitted over 10 km standard single mode fiber (SSMF) to the remote radio head (RRH) antenna site. The photo-mixing of these three components on the WIN-PDA, at a remote unit antenna site, results in the generation of sub-THz data signal at 200 GHz ($F_S$) and sub-THz carrier at 195 GHz ($F_c$).

The WIN-PDA THz transmitter consists of an InP-based photodiode, operating in the optical C-band, and a monolithically attached bowtie antenna [10]. To enable an efficient opto-electronic conversion at (sub-)THz frequencies, the footprint of the photodiode is kept small (<50 μm) and the intrinsic absorber layer is less than 300 nm. Evanescent coupling from optical waveguide to the absorber resulted in >0.3 A/W responsivity despite the reduced absorber size. The PD is merged into the feeding point of a bowtie antenna with 400 μm overall length [10]. This WIN-PDA transmitter is designed for CW radiation generation up to 4.5 THz [17] in a spectroscopic scenario. However, its emitted power reduces to 10 μW at 500 GHz and 1.9 μW at 1 THz from the peak value of 343 μW at 115 GHz for CW radiation. In our experiment, the WIN-PDA generated the 200 GHz sub-THz signal that was transmitted to the receiver.

At the receiver a horn antenna captured the signal which was fed to a WR 4.3 waveguide-based mixer which down-converted the sub-THz signal to IF with a 195 GHz LO. The LO carrier was generated using a 6-time frequency multiplier and a 32.5 GHz RF source as shown in the receiver box in Fig. 1. The use of WR 4.3 waveguide limited the frequency of operation in the 170 – 260 GHz band. The IF signal was captured using a real-time scope (RTS) and offline processing was done using Matlab to compensate for the FO and PN and then analyze the signal's error vector magnitude (EVM) performance. No amplifier or collimator lenses were used in the sub-THz link – limiting the wireless transmission distance to 20 cm.

Subsequently, the frequency of generated sub-THz signal was varied from 170 to 260 GHz by tuning the second laser as shown in the combined signal's spectrum in Fig. 2. The frequency of the LO at the receiver was changed accordingly to down-convert the signal close to 5 GHz IF for all the cases. The spectrum of captured 4 GHz BW IF signal is shown in Fig. 1(i).

### 3. Results and Discussion

Firstly, the results of impairment compensation are discussed for the 200 MHz BW 200 GHz sub-THz signal case and then the system's performance is presented. The frequency drift of the free-running lasers results in offset with respect to the desired sub-THz frequency as discussed in the introduction section. The fiber lasers used here exhibit ~30 MHz frequency fluctuations and have ~1 kHz linewidth as shown in our previous work in [18]. The demodulation of the RTS captured IF signal at the receiver, without any digital processing, results in a distorted constellation as shown in Fig. 3(i)-(a). The frequency fluctuations result in loss of orthogonality of the OFDM signal and it is important to compensate it to achieve successful system implementation. The FO estimation Schmidl & Cox algorithm [12] used here first detects a special preamble symbol with two identical halves in the received signal and then performs correlation to estimate the FO. A correction was applied to the signal in order to reverse the FO. The constellation of the FO compensated signal is shown in Fig. 3(i)-(b). The rotated constellation points indicate the presence of phase noise, which contains a contribution from both the laser's phase noise and frequency multiplier used in the sub-THz receiver. Pilot tones from OFDM subcarriers were used for estimating this phase rotation and subsequent correction was applied. The constellation of the compensated signal is shown in Fig. 2(i)-(c). The excellent constellation, with 2.7% EVM, of 2 MHz SC spacing 200 MHz BW 200 GHz sub-THz OFDM signal shows the successful compensation of FO and PN impairments and implementation of the low SC spacing link. In the previous demonstration with similar DSP [19], we demonstrated the transmission of 125 kHz and higher SC spacing signals at 60 GHz.

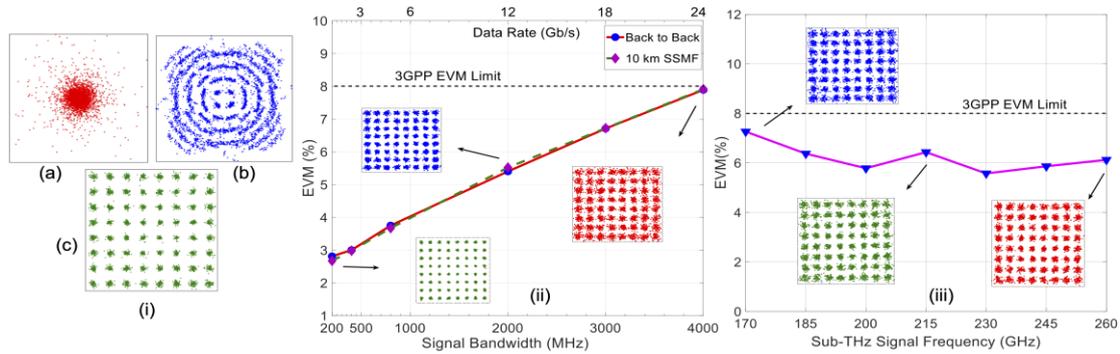

Fig. 3: (i) Constellation of the demodulated 200 MHz BW 200 GHz sub-THz signal at different stages of digital processing; (ii) EVM vs OFDM signal BW performance for 200 GHz sub-THz signal and (iii) EVM vs sub-THz frequency performance of 2 GHz bandwidth OFDM signal over 10 km optical heterodyne A-RoF link with WIN-PDA transmitter and sub-THz receiver.

In the second part, the BW of the OFDM signal was varied from 200 MHz to 4 GHz and the system's performance was analyzed at 200 GHz sub-THz frequency. The EVM curve, shown in Fig. 3(ii), indicates performance degradation with an increase in signal BW. This degradation stems from reduced SNR as the modulating signal power fed to IQ-MZM was constant for all the bandwidth values. The measured EVM of 7.89% and BER of ~2.2×$10^{-3}$ for 4 GHz was below the 8% 3rd generation partnership project (3GPP) EVM limit– showing the successful generation and demodulation of 24 Gb/s data at 200 GHz. Achieving high data rates with the WIN-PDA THz transmitter designed for CW operation is encouraging and shows the capability of the link for deployment in 6G systems.

In the third part, we analyzed the system's performance for 2 GHz BW OFDM signal at different sub-THz frequencies from 170 to 260 GHz. The frequency range limitation stems from the use of a WR 4.3 waveguide in the sub-THz mixer. The EVM results in Fig. 3(iii) show less than 1.75 % EVM variation as the sub-THz signal's frequency is changed. This variation comes from the non-ideal frequency response of the various components used in the system. In all the cases, the performance of the signals is below the 3GPP EVM limit - showing the flexibility and robustness of our system to generate variable frequency sub-THz signals.

4. **Conclusion**

An optical heterodyne A-RoF link provides an efficient solution for the generation and distribution of sub-THz/THz signals for the 6G system. In order to facilitate the wide deployment of such fiber-wireless systems, efficient use of multiple components such as widely tunable lasers, high-efficiency photodiodes, transmitter antennas and robust digital impairment compensation techniques is paramount. The results presented here show the capabilities of 'off-the-shelf' fiber lasers, a high-speed THz PDA and established DSP for the successful generation and transmission of low subcarrier spacing 160 – 270 GHz sub-THz OFDM signals. The flexibility and robustness offered by these components for the transmission of low subcarrier spacing OFDM signals pave the way for their deployment in the 6G system.

5. **Acknowledgements**

This work is funded by SFI grants: 18/SIRG/5579, 13/RC/2077_p2, 12/RC/2276 and ESA contracts 4000138240/22/NL/GLC/my, 4000139033/22/NL/IB/gg.